\documentclass[11pt,dvips]{article}
\usepackage{epsfig,times} 
\usepackage{picinpar}
\usepackage{wrapfig}
\usepackage{floatflt}
\makeatletter
\renewcommand{\section}{\@startsection{section}{1}{0in}
	{0.4\baselineskip}{0.1\baselineskip}{\Large\bf}}
\renewcommand{\subsection}{\@startsection{subsection}{2}{0in}
	{0.25\baselineskip}{-\baselineskip}{\large\bf}}
\renewcommand{\subsubsection}{\@startsection{subsubsection}{3}{0in}
	{0.1\baselineskip}{-\baselineskip}{\normalsize\bf}}
\makeatother
\begin{document}

%
\thispagestyle{myheadings}
\markright{HE 1.2.20}
\begin{center}
{\LARGE \bf Leading nucleon and the hadronic flux in the atmosphere}
\end{center}

\begin{center}
{\bf J. Bellandi$^{1}$, J.R. Fleitas$^{1}$, and J. Dias de Deus$^{2}$}\\
{\it $^{1}$Instituto de F\'{\i}sica, Universidade Estadual de Campinas, Campinas, SP 13083-970, Brazil\\
$^{2}$Instituto Superior T\'ecnico - CENTRA, Av. Rovisco Paes, 1, 1096 Lisboa Codex, Portugal}
\end{center}

\begin{center}
{\large \bf Abstract\\}
\end{center}
\vspace{-0.5ex}
Using an Iterative Leading Particle Model to describing nucleon-air collisions, we determine the hadronic flux in the atmosphere and compare with cosmic ray experimental data.
%

\vspace{1ex}

We present in this paper a calculation of the hadronic flux in the
atmosphere. Analytical solutions for the nucleonic diffusion equation in the
atmosphere having as boundary condition the primary spectrum and calculated
with the leading particle model, shows a strong correlation between the
inelastic proton-air cross section and the momentum of the leading particle
distribution, because the nucleonic shower penetration in the atmosphere
depends on the primary spectrum. In order to analyse experimental data on
the hadronic flux we correlate the moment of the leading particle
distribution in nucleon-air collisions with the respective one in
proton-nucleus coliisions using the Glauber model (Glauber 1959; Glauber et al. 1970).

For proton-nucleus scattering, at low energy, several models for describing
the leading particle spectrum have been proposed (Iteracting Gluon model
and Regge-Mueller formalism) (Yama et al. 1997; Dur\~aes et al. 1993; Batista et al. 1998) Here, we shall work in the Iterative
Leading Particle Model (Hwa 1984; Hufner \& Klar 1984) and use the notation of Frichter, Gaisser and
Stanev (Fritcher et al. 1997). In this model the leading particle spectrum in $
p+A\rightarrow N${\it (nucleon)}$+$ $X$ collisions is built from sucessive
interacions with $\nu $ interacting proton of the nucleus $A$ and the
behaviour is controlled by a straightforward convolution equation. It should
be mentioned that, strictly speaking, the convolution should be
3-dimensional. Here we only considered the 1-dimension approximation.

It is straightforward to show that in this model the $\gamma -th$ moment of the nucleon-air leading particle distribution, $<x^\gamma >_{N-air}$, is correlated wiht the respective nucleon-proton moment, $<x^\gamma >_{N-p}$,
by means of the following relation
\begin{eqnarray}
\sigma _{in}^{N-ar}(1-<x^\gamma >_{N-air})&=&
\int d^2b[ 1-[\eta (\gamma )\exp [-(1-K(\gamma )\sigma
_{tot}^{pp}T(b)]+ \nonumber \\
& & (1-\eta (\gamma )\exp [-\sigma _{tot}^{pp}T(b)]]] 
\label{1}
\end{eqnarray}
where $T(b)$ is the nuclear thickness and given by means of the Woods-Saxon model (Woods \& Saxon 1954; Barrett \& Jackson 1977)
\begin{equation}
\label{2}
\eta (\gamma )=\frac{<x^\gamma >_1^N}{K(\gamma )} 
\end{equation}
\begin{equation}
\label{3}
K_{\nu -1}(\gamma )=\beta _{\nu -1}\int_o^1dyy^\gamma S_\nu
^{+}(y)+(1-\beta _{\nu -1})\int_o^1dyy^\gamma S_\nu ^{-}(y) 
\end{equation}

The differential nucleonic flux in the atmosphere at some depth $t$ ,is
given by 
\begin{equation}
\label{4}
F_N(E,t)=N_oE^{-(\gamma +1)}\exp [-\frac t{\Lambda (E)}] 
\end{equation}
where $N_oE^{-(\gamma +1)}$ is the primary spectrum at $t=0$ and $\Lambda
(E) $ is the nucleonic attenuation length wich is given by 
\begin{equation}
\label{5}
\frac 1{\Lambda (E)}=\frac{\sigma _{in}^{N-ar}(1-<x^\gamma
>_{N-air})}{24100}(g/cm^2)^{-1} 
\end{equation}

We tested Eq. (\ref{1}) in comparison with cosmic ray data on nucleonic flux  (Brook et al. 1964; Ashton et al. 1968; Ashton et al. 1970)
and hadronic flux in the atmosphere (Mielke et al. 1993; Mielke et al. 1994; Aglietta et al. 1997). For $\sigma _{tot}^{pp}$ we
used the UA4/2 Collaboration parametrization (Burnett et al. 1992) and estimated $\sigma
_{in}^{N-air}$ by means of the Glauber model (Glauber 1959; Glauber et al. 1970) and for the $T(b)$ nuclear
thickness we used the Woods-Saxon model  (Woods \& Saxon 1954; Barrett \& Jackson 1977). The parameters $\eta $ and $K$
were left free. The best fit $(\aleph ^2/d.o.f.=2.61)$ corresponds to $\eta =1$
and $K=0.34$ and is shown in Fig. 1 (nucleonic flux at sea level) (Brook et al. 1964; Ashton et al. 1968; Ashton et al. 1970)  in
Fig. 2 (hadronic flux at sea level) (Mielke et al. 1993; Mielke et al. 1994) and in Fig. 3 (hadronic flux at $%
t=840g/cm^2$) (Aglietta et al. 1997). The hadronic flux was obtained multiplying the nucleonic
flux in Eq. (\ref{6}) by the Kascade factor (Mielke et al. 1993; Mielke et al. 1994)
\begin{equation}
\label{6}
R=\frac{\pi ^{+}+\pi ^{-}}{p+n}=0.04+0.27\ln (E/GeV) 
\end{equation}
in order to count the number of pions in the hadronic flux. We have used the
same primary spectrum as in (Bellandi et al. 1998). We also show in these figures the
maximal and minimal values for the flux considering the experimental errors
in the primary spectrum.

\vspace{1.ex}

We would like to thank the Brazilian governmental agencies CNPq and CAPES
for financial support.

\vspace{-0.8cm}
\begin{figure}
\begin{center}
{\mbox{\epsfig{file=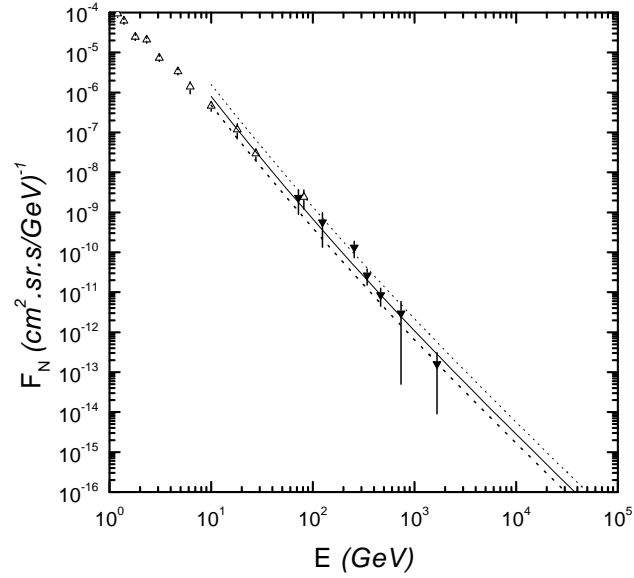,width=3.8in}}}
\end{center}
\vspace{-1.cm}
\caption{Nucleonic flux at sea level. Experimental data from (Brook et al. 1964; Ashton et al. 1968; Ashton et al. 1970). Continuous line, result of fit. Dash lines, maximal and minimal values of the calculated nucleonic flux.}
\end{figure}
\begin{figure}
\begin{center}
{\mbox{\epsfig{file=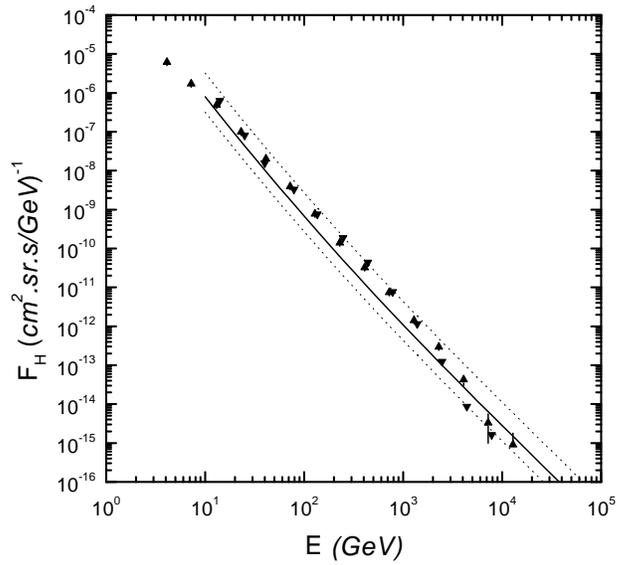,width=3.7in}}}
\end{center}
\vspace{-1.cm}
\caption{Hadronic flux at sea level. Experimental data from (Mielke et al. 1993; Mielke et al. 1994). Continuous line, result of fit. Dash lines, maximal and minimal values of the calculated hadronic flux.}
\end{figure}
\begin{figure}
\begin{center}
{\mbox{\epsfig{file=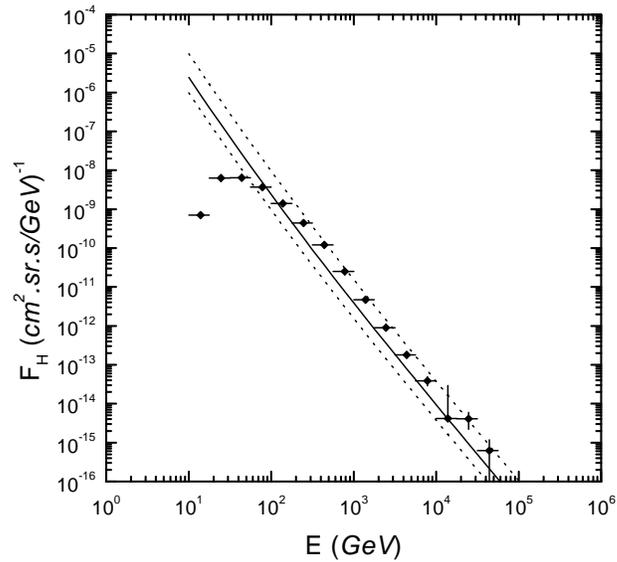,width=3.7in}}}
\end{center}
\vspace{-1.cm}
\caption{Hadronic flux at $t=830$ $g/cm^2$. Experimental data from (Aglietta et al. 1997). Continuous line, result of fit. Dash lines, maximal and minimal values of the calculated hadronic flux.}
\end{figure}

\newpage

\vspace{1ex}
\begin{center}
{\Large\bf References}
\end{center}
Aglietta, M. {\it et al.} (EASTOP Collab.) 1997 - Proc. 25$^{\rm th}$ ICRC (Durban, 1997), Vol.6, 81. \\
Ashton, F. {\it et al.} 1969 - J. Phys. {\bf A 1}, 169. \\ 
Ashton, F. {\it et al.} 1970 - Acta Phys. Acad. Sci. Hung. (Suppl 3) {\bf 29}, 25. \\
Barrett, R.C., \& Jackson, D.F. 1977 - Nuclear Sizes and Structure, (Clarendon Press - Oxford). \\
Batista, M. \& Covolan, R.J.M. 1998 - hep-ph:9811425. \\
Bellandi, J. {\it et al.} 1998 - Nuovo Cimento {\bf A111}, 149. \\
Brook, G. {\it et al.} 1964 -  Proc. Phys. Soc.{\bf 83}, 843. \\
Burnett, T.H. {\it et al.} 1992 - ApJ {\bf 349}, L25. \\
Dur\~aes, F.O. {\it et al.} 1993 - Phys. Rev. {\bf D 47}, 3049. \\
Frichter, G.M. {\it et al.} 1997 - Phys. Rev. {\bf D 56}, 3135. \\
Glauber, R.J. 1959 - Lect. Theor. Phys. Vol.1, edited by W.Britten and L.G.Dunhan (Interscience, NY), 135. \\
Glauber, R.J. {\it et al}. 1970, Nucl. Phys.{\bf B 12}, 135. \\
Hama, Y. {\it et al.} 1997 - Phys. Rev. Lett. {\bf 78}, 3070. \\
Hufner, J., \&  Klar, A. 1984 - Phys. Let {\bf 246B}, 167. \\
Hwa, R.C. 1984 - Phys. Rev. Lett. {\bf 52}, 492. \\
Mielke, H.M. {\it et al.} 1994 - J. Phys. {\bf G 20}, 637. \\ 
Mielke, H.M. {\it et al.} 1993 - Proc. 23$^{\rm th}$ ICRC, (Calgary, 1993) 4, 155. \\
Woods, R.D. \&  Saxon, D.S. 1954 - Phys. Rev. {\bf 95}, 577. \\

\vspace{1cm}

\noindent {\it Contribution to the 26$^{\rm th}$ International Cosmic Ray Conference, Salt Lake City, Utah, August 1999.}

\end{document}